%% file: main.tex
\pdfoutput=1
\documentclass[manuscript]{acmart}

\usepackage[utf8]{inputenc} 
\usepackage{algorithm} 
\usepackage{algorithmic} 
\usepackage{amsmath}
\usepackage{textcomp} 
\usepackage{subcaption} 
\usepackage{wrapfig} 
\usepackage{amsthm} 
\usepackage{multirow} 
\usepackage{mathtools} 
\usepackage{amsfonts} 
\usepackage{tabularx}
\usepackage{makecell}
\usepackage{array}
\usepackage{booktabs}

\keywords{algorithmic fairness, misinformation detection, large language models, generative AI, socio-technical systems}

\title{Diverse, but Divisive: LLMs Can Exaggerate Gender Differences in Opinion Related to Harms of Misinformation}

\begin{document}

\author{Terrence Neumann}
\affiliation{%
\institution{University of Texas at Austin}
\streetaddress{2110 Speedway}
\city{Austin}
\state{TX}
\country{USA}
\postcode{78705}
}

\author{Sooyong Lee}
\affiliation{%
\institution{University of Texas at Austin}
\streetaddress{2110 Speedway}
\city{Austin}
\state{TX}
\country{USA}
\postcode{78705}
}

\author{Maria De-Arteaga}
\affiliation{%
\institution{University of Texas at Austin}
\streetaddress{2110 Speedway}
\city{Austin}
\state{TX}
\country{USA}
\postcode{78705}
}

\author{Sina Fazelpour}
\affiliation{%
\institution{Northeastern University}
\streetaddress{2110 Speedway}
\city{Austin}
\state{TX}
\country{USA}
\postcode{78705}
}

\author{Matthew Lease}
\affiliation{%
\institution{University of Texas at Austin}
\streetaddress{2110 Speedway}
\city{Austin}
\state{TX}
\country{USA}
\postcode{78705}
}

\setcopyright{none}

\input{abstract}

\maketitle

\section{Introduction}
\input{introduction}

\section{Background}
\input{relatedwork}

\section{Approach}
\label{sec:approach}
\input{approach}

\section{Experiments}
\label{sec:experiment}
\input{experiments}

\section{Discussion}
\input{discussion}

\bibliographystyle{ACM-Reference-Format}
\bibliography{biblio}

\clearpage
\section{Appendix}
\input{appendix}

\end{document}

%% file: abstract.tex
\begin{abstract}
The pervasive spread of misinformation and disinformation poses a significant threat to society. Professional fact-checkers play a key role in addressing this threat, but the vast scale of the problem forces them to prioritize their limited resources. This prioritization may consider a range of factors, such as varying risks of harm posed to specific groups of people. In this work, we investigate potential implications of using a large language model (LLM) to facilitate such prioritization. Because fact-checking impacts a wide range of diverse segments of society, it is important that diverse views are represented in the claim prioritization process. This paper examines whether a LLM can reflect the views of various groups when assessing the harms of misinformation, focusing on gender as a primary variable. We pose two central questions: (1) To what extent do prompts with explicit gender references reflect gender differences in opinion in the United States on topics of social relevance? and (2) To what extent do gender-neutral prompts align with gendered viewpoints on those topics? To analyze these questions, we present the \texttt{TopicMisinfo} dataset, containing 160 fact-checked claims from diverse topics, supplemented by nearly 1600 human annotations with subjective perceptions and annotator demographics. Analyzing responses to gender-specific and neutral prompts, we find that \texttt{GPT 3.5-Turbo} reflects empirically observed gender differences in opinion but amplifies the extent of these differences. These findings illuminate AI’s complex role in moderating online communication, with implications for fact-checkers, algorithm designers, and the use of crowd-workers as annotators. We also release the \texttt{TopicMisinfo} dataset to support continuing research in the community. 
\end{abstract}

%% file: introduction.tex
The proliferation of information, including strategic disinformation and mistaken misinformation, challenge the integrity of public discourse and democracies worldwide~\cite{eliassi2020science,del2016spreading}. Professional fact-checkers play a critical societal role in addressing this challenge, but the scale of the problem demands some degree of automation to assist and accelerate the manual work required. At the same time, the complexity of fact-checking in the wild remains largely beyond the capabilities of even state-of-the-art AI  \cite{nadeem2019fakta, nakov2021automated, das2023state}. This recommends a hybrid human-AI teaming approach \cite{national2021human}, also referred to as mixed-initiative \cite{horvitz1999principles} or human-in-the-loop \cite{demartini2020human}, for technology-assisted fact-checking. 

A crucial step at the outset of fact-checking is \emph{claim prioritization} \cite{sehat2023misinformation}: how a fact-checker or their organization decides which {\em claims} (supposed facts) to check given limited resources.  Because far more claims appear online than professional fact-checkers could ever review, prioritization is crucial in allocating limited fact-checker resources to achieve the greatest impact. Such prioritization typically incorporates a range of factors~\cite{alam2021fighting}, of which a key consideration is the potential for harm \cite{sehat2023misinformation}. To assist such prioritization, a variety of machine learning techniques have been proposed, as well as community benchmarks, for inferring the \emph{check-worthiness} of claims \cite{nakov2022overview, jaradat2018claimrank, hassan2017toward}. With the emergence of Large Language Models (LLMs) and their increasing use across natural language processing (NLP) tasks, it seems likely that LLMs will be increasingly applied to fact-checking tasks, including check-worthiness prediction \cite{moorish2023fact, abels2023chatgpt}.

In this paper, we investigate potential implications of using LLMs for claim prioritization. The decision-making process in prioritization is complex, involving various factors that must be weighed relative to one another in making tradeoffs. It is also dynamic, adapting to the ever-changing landscape of information. Moreover, the choices made at this juncture can have far-reaching effects on different groups within society \cite{neumann2022justice, neumann2023does, sehat2023misinformation, alam2021fighting}. For example, prior to 2023, fact-checking organization Snopes claimed to ``write about whatever items the greatest number of readers are asking about'' \cite{snopes22}, a simple form of utilitarianism serving majority interests while potentially neglecting minority harms. Such an oversight can perpetuate the spread of misinformation within or impacting marginalized communities, which could lead to real-world harms and further exacerbate social disparities. 

Even apart from the equity of these downstream outcomes, critical issues of justice and fairness arise with respect to the characteristics of the allocation \textit{process} itself. Chief among these characteristics is the possibility of effective \textit{participation} in shaping the process. Specifically, a prioritization process that undermines the participation of particular groups can be seen as procedurally unfair (e.g., by unjustifiably disregarding or else discounting their particular concerns about what urgently needs fact-checking) ~\cite{herian2012public,zhang2015perceived}. In addition, such disregard for the views of a particular group can result in algorithmic tools whose outputs potentially harm members of that group~\citep{davani2022dealing}. 

To better understand the ethical implications of the use of LLMs, we examine whether LLMs can reflect the concerns and priorities of various groups when assessing the harms of misinformation, focusing on gender as a primary variable. We examine gender as a variable of interest due to its established influence on perspective as documented in social science research \cite{lizotte2020gender, huddy2008gender, ANES2021}; this relevance is further underscored by the observed differences in how misinformation may impact or target genders differently \cite{thakur_hankerson_2021}. Our dataset, \texttt{TopicMisinfo}, includes human annotations that align with the prior research, providing an empirical basis for our analysis. This dataset is a collection of 160 recently fact-checked claims sourced from the Google Fact-Check API covering a wide range of topics. It is supplemented by 1592 human annotations, which provide insights into their perceptions across a variety of prioritization criteria and demographics of the annotators.

This novel dataset allows us to investigate two central questions: (1) To what extent do prompts with explicit gender references reflect gender differences in opinion in the United States on topics of social relevance? and (2) To what extent do gender-neutral prompts align with gendered viewpoints on those topics?

Our analysis includes a detailed examination of a widely adopted LLM's (\texttt{GPT-3.5 Turbo}) responses to both gender-conditioned and gender-neutral prompts. We assess the degree to which these AI models are able to reflect empirically observed gendered opinion differences. Interestingly, our findings reveal a complex picture: while LLMs demonstrate a capability to mirror certain gender-specific opinion trends, they also show a tendency to amplify gender discordance when responding to topics where both prior work and our empirical data of human annotations suggests there should be very little. We also find that, when prompted without gender-conditioning, \texttt{GPT-3.5 Turbo} demonstrates more alignment with men when assessing the potential harm to groups, particularly when assessing claims on divisive topics. These results highlight the challenges inherent in using AI to reflect human opinions accurately and fairly.

The paper is structured as follows: Section 2 delves into recent studies on LLMs simulating human behavior and provides background about the fact-checking pipeline, automation, and ethical challenges therein. In Section 3, we present our approach to data collection. We also describe our technique for prompting LLMs to elicit their responses to a claim prioritization task. Section 4 highlights our methodology and presents our findings. Finally, Section 5 examines the broader implications of LLMs in claim prioritization for various stakeholders, including fact-checkers, LLM developers, and crowd-workers. We hope that this research, along with the release of the \texttt{TopicMisinfo} dataset, contributes to the ongoing dialogue and development in the field, encouraging further research and innovation in responsible AI-driven fact-checking tools.

%% file: relatedwork.tex
\subsection{LLMs for Subjective Annotation Tasks}

There is substantial interest in exploring the ability of Large Language Models (LLMs) to emulate human subjects \cite{aher2023using, argyle2023out, horton2023large} and the feasibility of using LLMs to generate subjective survey results or training labels, to potentially replace human input \cite{zhu2023can}. These works are driven by the hope that, if these models could represent the viewpoints of various sub-populations (a concept known as algorithmic fidelity), they could significantly enhance our understanding of human social behaviors, while offering the benefits of speed and cost efficiency \cite{korinek2023language}.

Understanding the perspectives LLMs acquire during their training is crucial when considering them as surrogates for human opinions. Research has investigated the alignment of LLM opinions with those of diverse human populations \cite{santurkar2023whose, perez2022discovering, argyle2023out}. Findings indicate that the viewpoints expressed by LLMs vary based on their training. For example, models trained with Reinforcement Learning from Human Feedback (RLHF) tend to demonstrate strong political biases, often skewing more towards liberal rather than conservative stances on issues like immigration and gun rights \cite{perez2022discovering}. While one study suggests GPT-3 (davinci), one popularly used LLM, shows high algorithmic fidelity in mirroring public opinion across U.S. sub-populations \cite{argyle2023out}, others argue the opposite \cite{santurkar2023whose}. For instance, text-davinci-003 seems to reduce opinion diversity by assigning extremely high probabilities to certain choices in multiple-choice questions, indicating a lack of nuanced representation of different group opinions \cite{santurkar2023whose}.

Another aspect of this research involves examining whether language models can be prompted to reflect the opinions of specific demographic groups. Studies have shown mixed results \cite{santurkar2023whose, sun2023aligning, aher2023using, horton2023large}; while some LLMs from OpenAI and AI21 Labs become more representative of certain sub-populations after prompting \cite{santurkar2023whose, aher2023using}, experiments using newer OpenAI models (GPT-3.5 and GPT-4) to predict group perspectives on politeness and offensiveness indicate a contrary trend \cite{sun2023aligning}. Notably, including a target demographic in the prompt often reduces model accuracy, with this effect being particularly pronounced when simulating opinions of Asian and Black racial groups. \citet{agnew2024illusion} describe the process of using LLMs to capture diverse participant viewpoints as ``illusory'' as it does not incorporate the fundamental values of human-subject research, such as representation, inclusion, and understanding.

\subsection{The Fact-Checking Pipeline, Automation, and Fairness}  

Fact-checking is a multi-step process crucial for distinguishing truth from falsehood in the digital information landscape. As the volume of misinformation grows, there is a growing interest in leveraging AI to enhance the efficiency and scalability of the fact-checking pipeline \cite{das2023state, guo2022survey, zeng2021automated, nakov2021automated}. There are now annual conferences and workshops which focus on developing technologies to assist fact-checkers in these tasks \cite{CheckThat, liu2023human}.

The fact-checking pipeline begins with \emph{claim detection}, identifying verifiable claims within trending news items, and distinguishing them from opinions or personal thoughts \cite{das2023state, guo2022survey}. This is followed by \emph{claim matching}, which checks whether the claim has already been investigated to prevent redundant efforts \cite{das2023state}. Social media companies, recognizing the importance of this step, have invested in technologies to halt the spread of already debunked misinformation \cite{meta_ai_2020}.

The third phase, \emph{claim prioritization}, assesses the importance of investigating a claim based on various factors, including potential harm \cite{nakov2021automated, alam2021fighting}. This step is crucial, as it determines which claims are deemed worthy of verification. Following this, \emph{evidence retrieval} involves sourcing evidence to support or contradict the claim \cite{nadeem2019fakta, samarinas2021improving}. The final stage, \emph{verdict prediction}, uses this evidence to conclude the claim’s validity. However, automating this stage remains contentious due to the scrutiny required, a level AI cannot currently match. As \citet[p.3]{nakov2021automated} note, incorrect automated fact-checks could severely damage a fact-checking organization’s reputation.

Our study focuses on claim prioritization, a contentious and multi-faceted task with significant implications for the entire fact-checking process. AI's potential in enhancing efficiency here is noteworthy; for instance, ClaimHunter, an AI-based claim prioritization tool, reportedly improved claim selection efficiency at a fact-checking organization by 70-80\% \cite{moorish2023fact}. Fact-checking prioritization, as \citet{sehat2023misinformation} outlines, involves considering multiple dimensions: Fragmentation, Actionability, Believability, Likelihood of Spread, and Exploitativeness. Misinformation that targets specific vulnerable groups tends to be prioritized along several of these dimensions.

However, concerns arise regarding AI systems' potential biases in this prioritization process. \citet{neumann2022justice} cautions about the adverse effects of such biases, particularly harming groups either misrepresented in unchecked misinformation or seeking accurate information on sensitive topics. Furthermore, \citet{neumann2023does} warn that even seemingly minor design choices in AI systems can lead to significant disparities in outcomes, disproportionately affecting underrepresented groups.

Understanding the fairness of AI used in different stages of fact-checking pipelines could be complicated due to a variety of reasons, such as the lack of an independently agreed upon standard for the ethically desirable outcomes. Furthermore, even if we had such a standard, the many intermediary steps in the fact-checking pipeline coupled with complex social dynamics further complicates one's ability to connect choices at the prioritization stage to downstream societal outcomes. In such scenarios, the fairness of outcomes (what gets prioritized for fact-checking and what does not) might not be the most appropriate focus. Instead, ensuring a fair process in obtaining fact-checks might be more crucial \cite{herian2012public, zhang2015perceived}. This perspective is particularly important when considering replacing human annotators with LLMs in the claim prioritization task. It is essential to evaluate whether these models can accurately reflect the prioritization judgments of diverse groups, especially when these groups have different concerns about certain topics. In what follows, we delve deeper into these concerns and discuss their implications.

%% file: approach.tex
\subsection{Theoretical Grounding}

In exploring public opinion and subjective data annotation, it is important to recognize how societal factors, including gender, can influence individual perspectives and values. However, it is crucial to avoid essentialist views that inherently attribute certain values or opinions to a gender. Rather, differences observed in public opinion across genders are reflective of aggregate trends at specific times and places, shaped by the varying social environments and experiences of individuals \cite{intemann201025}.

Research, such as that by \citet{lizotte2020gender} and \citet{huddy2008gender}, indicates that, in certain contexts, women as a group have shown a tendency to prioritize issues like equality and social welfare more than men. For instance, Lizotte's mediation analysis reveals that values such as universalism, benevolence, and egalitarianism are adopted and prioritized differently by women and men, influencing their perspectives on various issues, including social justice and welfare policies \cite{lizotte2020gender}. Data from the 2020 American National Election Studies (ANES) survey also reflects these differences in opinion \cite{ANES2021}. It shows that, on average, women in the surveyed population expressed more egalitarian attitudes on issues like LGBTQ rights, immigration, racial equality, and reproductive rights (see Table \ref{tab:poll_data} in the Appendix). Note that the differences observed here are relatively small, but statistically significant.

Our study design builds on this prior work and is informed by the expectation of finding gender differences in opinion on certain topics but not on others. This approach allows us to investigate whether and how these differences manifest in the context of identifying harms of potential misinformation. For example, if women generally exhibit a stronger inclination towards values like universalism and communal welfare, they might perceive misinformation threatening particular demographic groups as more harmful. In contrast, men, who may prioritize values like autonomy, could be more concerned about misinformation impacting economic stability or personal freedoms. These hypotheses based on observed trends inform our data collection process and well as our experimental design.

\subsection{Data Collection}

In building the \texttt{TopicMisinfo} dataset, our aim was to compile crowd-workers' perceptions of the need for fact-checking of both true and false information assessed across multiple dimensions on a broad range of topics. This would enable us to effectively gauge how well LLMs reflect diverse opinions regarding the impact of misinformation across these varied subjects. Using the logic established in the previous section, we hypothesized that certain topics were likely to generate disagreement on the assessed potential harm dependent on the demographics of the workers annotating the claims. In particular, using insights from the ANES 2020 polling data \cite{ANES2021}, we thought that claims related to (i) reproductive rights and abortion, (ii) African American issues, (iii) immigration policy, and (iv) LGBTQ issues were particularly likely to generate disagreement on gender lines, based on the differing values of men and women. We also gathered claims from topics in which we didn't expect large gender disagreement when assessing potential harm, such as (i) health and science issues, (ii) claims about U.S. politicians, the U.S. military, and U.S. foreign policy (iii) weather and climate claims, (iv) claims about entertainers, and (v) sports-related claims. To source these claims, we employed Google’s Fact-Check API\footnote{\url{https://toolbox.google.com/factcheck/explorer}.}, filtering first by the specified topics, then again for articles written in English. This API consolidates recent fact-checks from reputable sources, including Snopes, PolitFact, and FullFact. Because the API includes the date of the fact-check, we were able to ensure that all claims in our dataset were recent. 

Using Amazon Mechanical Turk, we hired American workers and compensated them at a rate of \$15-20 per hour to annotate various claims. In particular, we asked them to assess four dimensions of each claim: (1) need for prioritization in fact-checking; (2) interest to the general public; (3) potential to harm certain demographic groups; and (4) perceived validity. For each assessment, we recorded the answers on a 1-to-6 Likert scale (see Table \ref{tab:definitions} in the Appendix for the verbatim criteria workers used to evaluate claims). If workers identified a claim as harmful to a specific group or groups in (3), we asked them to identify which group or groups may be harmed. Additionally, we had workers fill out a demographic survey in which they optionally provided their age range, gender identity, education level, and sexual orientation. Almost all workers voluntarily provided this information.

In total, we gathered 1998 annotations spanning 160 distinct claims. To ensure the quality and reliability of our data, we introduced attention checks in which we could assume there exists ``gold'', or objective, answers. These gold questions were either manifestly false (e.g.,``The largest tree creates a land bridge between the earth and Mars'') or patently true (e.g., ``A circle is round''). We retained only the data from workers who demonstrated over 80\% accuracy in their perception of the veracity of gold claims. Concurrently, we eliminated a small number of duplicate responses in which a worker evaluated the same claim multiple times. This filtering yields 1592 annotations from 27 distinct workers, which were then subjected to analysis. See Table \ref{tab:distributions} below and Table \ref{tab:demographics} in the Appendix for relevant metadata about the dataset. Notably, 69\% of our annotations came from workers identifying as men, 31\% came from workers identifying as women, and we were unable to obtain annotations from anyone identifying as non-binary.

\begin{table*}[h]
\centering
\caption{Distribution of TRUE and FALSE claims by topic, with corresponding number of annotations gathered by gender for each topic.}
\begin{tabular}{|l|c|c|c|c|}
\hline
\textbf{Topic} & \textbf{\# FALSE Claims} & \textbf{\# TRUE Claims} & \textbf{\# Men Annotations} & \textbf{\# Women Annotations}\\
\hline
Abortion & 13 & 6 & 119 & 61 \\
\hline
Black Americans & 6 & 6 & 78 & 37 \\
\hline
Entertainment & 5 & 4 & 63 & 23\\
\hline
Gold & 11 & 15 & 199 & 141 \\
\hline
Health and Science & 9 & 1 & 71 & 30\\
\hline
Illegal Immigration & 11 & 3 & 88 & 37\\
\hline
LGBTQ & 10 & 4 & 93 & 45 \\
\hline
Sports & 3 & 2 & 27 & 17 \\
\hline
USA & 38 & 5 & 280 & 120 \\
\hline
Weather and Climate & 3 & 5 & 42 & 21\\
\hline
\hline
\textbf{Total} & \textbf{109} & \textbf{51} & \textbf{1060} & \textbf{532} \\
\hline
\end{tabular}
\label{tab:distributions}
\end{table*}

\subsection{Prompt Design \& LLM Annotation Collection}

\subsubsection{Prompt Design}
We design our prompts based off of prior work (\cite{argyle2023out, sun2023aligning, zhu2023can}) and best practices suggested by OpenAI \footnote{https://platform.openai.com/docs/guides/prompt-engineering/six-strategies-for-getting-better-results}. We prompt \texttt{GPT-3.5 Turbo} with the same description of group harm that was presented to Turkers to acquire annotations:
\begin{center} \emph{How likely is this claim to disproportionately harm certain demographic groups more than others (e.g. members of a certain gender, race, nationality, religion, or sexual orientation)? (1-6 scale)}
\end{center}

For gendered-conditioned prompting, we considered two variants, included in Table \ref{tab:combinedprompts}. For prompt 1, we use the persona pattern \cite{white2023prompt} and ask the model to naturally complete the sentence in order to elicit its score prediction. We observe that prompting the model to complete the sentence and think as a gender circumvents safety mechanisms within \texttt{GPT-3.5 Turbo} which are common in other large language models \cite{wei2023jailbroken}. In order to avoid these safety mechanisms which can sporadically occur when prompting language models for opinionated responses, we mention the gender twice in the prompt.

For prompt 2, we follow the prompting strategies of \citet{sun2023aligning} and \citet{zhu2023can}. We similarly design prompt 2 with the aforementioned safety mechanisms in mind, as delineating specific options for the model to respond with (e.g. 1, 2, 3, 4, 5, 6) encourages the model to respond with only these answers. Additionally, prompting the model to provide the answer without explaining its reasoning results in succinct, shorter responses which reduces the tokens generated. This prompt design also minimizes the likelihood of the model to refuse giving its opinion as a result of its safety mechanisms. This refusal typically takes the form of "As an AI language model, I do not have personal opinions or biases." The gender-neutral prompts 1 and 2 are similar to the gender-conditioned prompts 1 and 2 without the additional gender context information embedded into the prompt.

\begin{table*}[h]
\centering
\begin{tabular}{|l|l|}
\hline
\multicolumn{2}{|c|}{\textbf{Gender-Conditioned Prompts}} \\
\hline
\textbf{Prompt} & \textbf{Format} \\
\hline
1 & \makecell[l]{Claim: "\{claim\}" \\Answer the following question as if you were a \{gender\}.\\How likely is this claim to disproportionately harm certain demographic groups more than others \\(e.g. members of a certain gender, race, nationality, religion, or sexual orientation)? (1-6 scale)\\As a \{gender\}, on a scale from 1 to 6 I would rate it as a} \\
\hline
2 & \makecell[l]{Answer the following question as if you were a \{gender\}.\\How likely is this claim to disproportionately harm certain demographic groups more than others \\(e.g. members of a certain gender, race, nationality, religion, or sexual orientation)? (1-6 scale)\\Options: 1, 2, 3, 4, 5, 6\\Claim: "\{claim\}"\\Response (Provide the answer without explaining your reasoning):} \\
\hline
\multicolumn{2}{|c|}{\textbf{Gender-Neutral Prompts}} \\
\hline
\textbf{Prompt} & \textbf{Format} \\
\hline
1 & \makecell[l]{Claim: "\{claim\}" \\How likely is this claim to disproportionately harm certain demographic groups more than others \\(e.g. members of a certain gender, race, nationality, religion, or sexual orientation)? (1-6 scale)\\On a scale from 1 to 6 I would rate it as a} \\
\hline
2 & \makecell[l]{How likely is this claim to disproportionately harm certain demographic groups more than others \\(e.g. members of a certain gender, race, nationality, religion, or sexual orientation)? (1-6 scale)\\Options: 1, 2, 3, 4, 5, 6\\Claim: "\{claim\}"\\Response (Provide the answer without explaining your reasoning):} \\
\hline
\end{tabular}
\caption{Prompts Used in LLM Experiments}
\label{tab:combinedprompts}
\end{table*}

\subsubsection{LLM Annotation Collection}

LLM annotations are derived from three distinct conditions: annotations prompted to reflect the views of men, the views of women, and prompted neutrally without specifying a gender, which we refer to as \emph{base} perspective. In each case, the set of LLM annotations has the same number of elements as the human annotations for a given claim. For example, if a claim has 7 annotations by women and 3 by men, we prompt the LLM 7 times with the women-conditioned prompt and 3 times with the men-conditioned prompt. When prompting for the base perspective, we gather the total number of human annotations for a given claim. Referring to the example above, we would gather 10 ratings using the base prompt. It is worth noting that we set the \emph{temperature} parameter to 0, so we get very little variance in the claim-level rating by the LLM. We repeat the annotation collection process for both prompts illustrated in Table \ref{tab:combinedprompts}.

%% file: experiments.tex
In the previous section, we described our dataset and methodology for data collection. Building on that foundation, we will now delve into our two primary research questions. To facilitate understanding, we first introduce some key terms and notation that will be consistently used in this discussion. Our dataset is structured hierarchically and encompasses a wide range of $D$ topics, collectively denoted as $\Omega = \{\omega_{1}, \omega_{2}, ..., \omega_{D}\}$. For simplicity and clarity in our discussion, we will refer to individual topics as $\omega$, dropping the subscript notation. Each topic $\omega$ has a set of claims associated to it, noted as $\{c_1, c_2, ..., c_{N_{\omega}}\}$ with $N_\omega$ elements. These claims can either be true or false. Importantly, every claim $c_{i}$ within a topic $\omega$ is evaluated by different \emph{sources}, which include both LLMs and human annotators. See the previous section for a description of our strategy in gathering annotations from LLMs and human annotators.

\subsection{RQ1: Do Gender-Conditioned Prompts Reflect Diversity of Opinion?}\label{sec:rq1}

\subsubsection{Methodology.} 

Our objective is to examine the extent to which patterns of gender disagreement observed in human responses are reflected in the responses generated by a LLM when prompts are gender-conditioned.  

To ensure an appropriate comparison of the ratings given by humans and LLMs, we need to consider potential differences in their scoring methods. While both sources were asked to use the same  Likert scale (1-6), it is possible that their usage of the scale differs. For example, LLMs might show a greater range in their scores, while humans may avoid using the extreme ends of the scale. If this were the case, noticeable differences in the scoring patterns of LLMs and humans might simply reflect their distinct approaches to using the scale, rather than a meaningful difference in their evaluations. To address this, we have standardized the scores. We have done this by applying a z-score transformation to the scores of each group, which normalizes the data and accounts for any differences in the use of the scale and the average scores between the two groups.

Once the scores are standardized, we can focus on assessing gender differences in opinion. We introduce a claim-level statistic to measure the gender differences in gender-prompted LLM responses, and compare these to the gender differences observed in human responses. Thus, for each claim $c_{i}$ $\in$ $\omega$:

\begin{equation}
\hat{E}_{c_{i}} = \frac{|\mu(c_{i})_{LLM}^{woman} - \mu(c_{i})_{LLM}^{man}| - |\mu(c_{i})_{Human}^{woman} - \mu(c_{i})_{Human}^{man}|}{\sqrt{1 + \sigma(c_{i})_{Human}^2 + \sigma(c_{i})_{LLM}^2}}
\end{equation}

In the above equation, $\mu(c_{i})$ represents the average z-score of the annotations from a specified source and gender-condition for a claim. For example, $\mu(c_{i})_{LLM}^{woman}$ is the mean z-score for woman-prompted LLM, and  $\mu(c_{i})_{Human}^{woman}$ is the mean z-score provided by annotators identifying as a woman. Correspondingly, $\sigma(c_{i})_{Human}^{2}$ and $\sigma(c_{i})_{LLM}^{2}$ represent the variance of z-scores of annotations for a claim $c_{i}$ for all human responses and LLM responses, respectively. The denominator of $\hat{E}_{c_{i}}$ helps control for claim-level variation in responses such that, when variation amongst either source is high, the value of the statistic is down-weighted.

To assess the overall trend within a topic, we aggregate this statistic to the topic level by calculating the average $\hat{E}_{c_{i}}$ for all $c_{i}$ in a topic $\omega$.

\begin{equation}
\hat{E}_{\omega} = \frac{1}{N_{\omega}}\sum_{i=1}^{N_{\omega}}\hat{E}_{c_{i}}
\end{equation}

Finally, we construct a hypothesis around the topic-level mean, $\hat{E}_{\omega}$, as follows:

\begin{equation}
\begin{array}{rcl}
H_0: \hat{E}_{\omega} = 0 \\
H_1: \hat{E}_{\omega} > 0 
\end{array}
\end{equation}

A value of $\hat{E}_{\omega} = 0$ implies that the LLM accurately reflects human disagreement levels across a topic $\omega$, whereas $\hat{E}_{\omega} > 0$ would indicate an exaggeration of disagreement amongst gender-conditioned responses by the LLM when compared to the observed human responses.

To reliably test our hypotheses, we use a bootstrap approach as recommended by \citet{hall1991two} and \citet{mackinnon2009bootstrap}. We assume $H_0$ is true; that is that there is no difference between LLM and human annotations for a given gender-condition. For each claim $c_{i}$, we sample with replacement the same number of annotations that appear for a given source. We repeat this process for each gender-condition in the the \texttt{TopicMisinfo} dataset. For instance, if a given claim had $x$ annotation from men, we would draw $x$ annotations with replacement from the combined pool of men's annotations and men-conditioned LLM annotations. From these bootstrapped samples of annotations, $\tilde{\mu}(c_{i})_{LLM}^{man}$, $\tilde{\mu}(c_{i})_{Human}^{man}$, $\tilde{\mu}(c_{i})_{LLM}^{woman}$, and $\tilde{\mu}(c_{i})_{Human}^{woman}$, along with their respective variance estimates $\tilde{\sigma}(c_{i})_{LLM}^{2}$ and $\tilde{\sigma}(c_{i})_{Human}^{2}$ are calculated. Thus our bootstrap statistic for a claim $c_{i}$ is:

\begin{equation}
    \tilde{E}_{c_{i}}^{b} = \frac{|\tilde{\mu}(c_{i})_{LLM}^{woman} - \tilde{\mu}(c_{i})_{LLM}^{man}| - |\tilde{\mu}(c_{i})_{Human}^{woman} - \tilde{\mu}(c_{i})_{Human}^{man}|}{\sqrt{1 + \tilde{\sigma}(c_{i})_{Human}^2 + \tilde{\sigma}(c_{i})_{LLM}^2}} 
\end{equation}

Each claim undergoes the bootstrap procedure $B$ times (with $B = 10000$ in our case). A topic-level bootstrapped mean is calculated for each bootstrap iteration $b$: 

\begin{equation}
\tilde{E}_{\omega}^{b} = \frac{1}{N_\omega}\sum_{i=1}^{N_\omega}\tilde{E}_{c_{i}}^{b}
\end{equation}

After all bootstrap iterations $B$ are complete, we then calculate $T$ as follows:

\begin{equation}
 T = \frac{1}{B} \sum_{b = 1}^{B} I[\tilde{E}_{\omega}^{b} > \hat{E}_{\omega}] 
\end{equation}

Here, $I[.]$ is the indicator function, which equals 1 if the enclosed condition is true and 0 otherwise. $T$ estimates the probability of observing a $\hat{E}_{\omega}$ at least as large as ours given that $H_0$ is true (i.e. $T$ estimates the p-value). We reject $H_0$ if $T < \alpha$, where $\alpha$ is the significance level. We repeat this process for LLM annotations gathered from Prompt 1 and Prompt 2 (see Table \ref{tab:combinedprompts} for details) and present the results in the next section.

\subsubsection{Results.}

\input{results_rq1}

\begin{figure*}
 \centering
    \includegraphics[width=\textwidth]{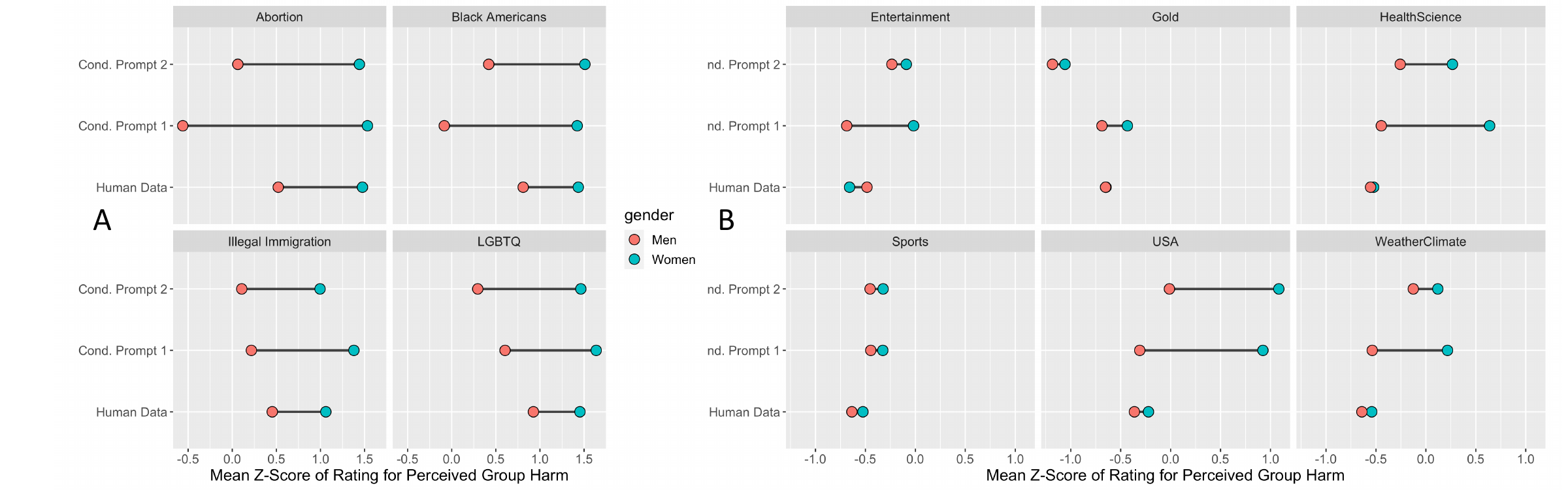}
    \caption{These charts, relevant to RQ1, illustrate variations in average responses when evaluating the perceived harm to specific demographic groups for claims across topics. Chart A compares the differences in opinions between human men and women with those generated by AI using gender-conditioned prompts on topics prone to divergent viewpoints. Here, we observe that prompts tend to capture and even exaggerate the range of opinions found in human responses. In contrast, Chart B reveals that AI models tend to forecast significant levels of disagreement on topics that typically do not cause such discordance in real-world scenarios, such as Health and Science and Weather and Climate.}
    \label{fig:rq1}
\end{figure*}

Table \ref{tab:results_rq1} presents the outcomes of our bootstrap test, while Figure \ref{fig:rq1} graphically delineates the gender differences and the corresponding responses from a gender-conditioned LLM prompt. The data shows a clear pattern: Prompt 1 tends to amplify the gender disparities in perception across a broad spectrum of topics. We see statistically significant positive values of \(\hat{E}_{\omega}\), which are observed across contentious topics such as Abortion (0.56), Black Americans (0.63), and Illegal Immigration (0.22), as well as more neutral topics like Entertainment (0.30) and WeatherClimate (0.47), all significant at a \(\alpha = 0.05\) significance level. Meanwhile, Prompt 2 echoes the trends of Prompt 1 but to a lesser degree. The lower \(\hat{E}_{\omega}\) values for Prompt 2, seen in subjects ranging from Abortion (0.24) to HealthScience (0.34), suggest that it more accurately mirrors the gender differences in opinion present in our dataset, except for LGBTQ issues where Prompt 2's \(\hat{E}_{\omega}\) is notably higher (0.31). 

The distinction in how each prompt addresses gender may account for the observed variations in responses. Specifically, Prompt 1 contains two distinct references to gender: initially before the rating prompt and subsequently within the response, as one of the last tokens. In contrast, Prompt 2 mentions gender only once at the beginning of the prompt. We hypothesize that this reinforced emphasis on gender in Prompt 1 is likely contributing to the more pronounced gender-based differences in ratings. Transformer models, the dominant architecture used to train language models, are autoregressive \cite{katharopoulos2020transformers}, meaning that they are sensitive to the ordering of input tokens. Therefore, when gender is highlighted more explicitly towards the end of a prompt, as seen in Prompt 1, this may increase the probability of replicating gendered stereotypes. Crucially, assessing which prompt better aligns with human assessments requires data collection of human responses, questioning the feasibility of using these tools to circumvent the need for human assessments.

\subsection{RQ2: Do Gender-Neutral Prompts Align with a Particular Group?}

\subsubsection{Methodology}

To investigate whether neutrally-prompted LLM responses are more aligned with the responses of men or women, we introduce a topic-level statistic that measures the difference in mean squared error between the neutrally-prompted LLM scores and the ratings of men and women. Similarly to RQ1, prior to calculation, we apply a z-score transformation to the human annotations and LLM annotations to account for possible differences in variance and scoring patterns. 

For each topic $\omega$, gender-level mean squared error ($MSE_{\omega}^{gender}$) is defined as:

\begin{equation}
    MSE_{\omega}^{gender} = \frac{1}{N_{\omega}}\sum_{i = 1}^{N_\omega}\left(\mu(c_{i})_{Human}^{gender} - \mu(c_{i})_{LLM}^{base}\right)^{2}
\end{equation}

where $\mu(c_{i})$ represents the average z-score of the annotations from a specified source for a claim. For each topic $\omega$, we define a topic-level statistic:

\begin{equation}
\hat{D}_\omega = |MSE_{\omega}^{woman} - MSE_{\omega}^{man}|
\end{equation}

and we construct a hypothesis around the topic-level statistic $\hat{D}_\omega$:

\begin{equation}
\begin{array}{rcl}
H_0: \hat{D}_\omega = 0 \\
H_1: \hat{D}_\omega \neq 0
\end{array}
\end{equation}

If $\hat{D}_\omega = 0$, this implies that the output scores from the neutrally-prompted LLM are equally distant from the responses of men and women. If $\hat{D}_\omega \neq 0$, then this implies that the LLM output is more aligned with one group over the other.

Again, similar to RQ1, we use bootstrapping to reliably test this hypothesis. We assume $H_0$ is true; that is, the $MSE_{\omega}^{woman}$ and $MSE_{\omega}^{man}$ are the same. In other words, $H_{0}$ assumes that the annotations of men and women come from the same distribution, and makes no assumption about the distribution of neutrally-prompted LLM responses. Thus, our bootstrapping procedure combines the observed annotations of men and women for $c_{i}$ and draws samples with replacement, replicating the number of annotations originally made by each gender for $c_{i}$. For instance, if claim $c_{i}$ has 7 annotations from men and 3 annotations from women, we would combine these 10 annotations and resample with replacement, yielding new sets of 7  and 3 bootstrapped annotations. We then take the mean value of these annotations ($\tilde{\mu}(c_i)_{Human}^{man}$ and $\tilde{\mu}(c_i)_{Human}^{woman}$) and compare them to the average of the neutrally-prompted LLM response ($\mu(c_{i})_{LLM}^{base}$).

\begin{equation}
        \tilde{MSE}_{\omega}^{gender} = \frac{1}{N_{\omega}}\sum_{i = 1}^{N_\omega}\left(\tilde{\mu}(c_{i})_{Human}^{gender} - \mu(c_{i})_{LLM}^{base}\right)^{2}
\end{equation}

Each topic undergoes this bootstrap procedure $B$ times (with $B = 10000$). For each claim $c_i$, we derive $\tilde{MSE}_{\omega}^{woman}$ and $\tilde{MSE}_{\omega}^{man}$ and define the topic-level bootstrapped statistic for each bootstrap iteration $b$ as:

\begin{equation}
    \tilde{D}_{\omega}^{b} = |\tilde{MSE}_{\omega}^{woman} - \tilde{MSE}_{\omega}^{man}|
\end{equation}

After all iterations $B$ are complete, we then calculate $T$ as follows:

\begin{equation}
 T = \frac{1}{B} \sum_{b = 1}^{B} I[\tilde{D}_{\omega}^{b} > \hat{D}_{\omega}]    
\end{equation}

As before, $I[.]$ is the indicator function. $T$ estimates the probability of observing a $\hat{D}_{\omega}$ at least as large as ours given that $H_0$ is true (i.e. $T$ estimates the p-value). We reject $H_0$ if $T < \alpha$, where $\alpha$ is the significance level.

\subsubsection{Results}

\begin{table*}[h]
\centering
\caption{Combined Results for Prompts 1 and 2}
\begin{tabular}{@{}lcccccc@{}} 
\toprule
Topic & $MSE_{men}$ (1) & $MSE_{women}$ (1) & p-value & $MSE_{men}$ (2) & $MSE_{women}$ (2) & p-value \\
\midrule
\multicolumn{7}{c}{Diverse Topics} \\
\midrule
Abortion & 0.86 & 2.16 & \textbf{0.00} & 0.57 & 1.69 & \textbf{0.00}\\
Black Americans & 0.74 & 0.53 & 0.48 & 0.78 & 0.78 & 0.99 \\
Illegal Immigration & 1.01 & 1.18 & 0.35 & 1.16 & 2.04 & \textbf{0.04}\\
LGBTQ & 1.02 & 1.27 & 0.53 & 1.00 & 1.26 & 0.47\\
\midrule
\multicolumn{7}{c}{Other Topics} \\
\midrule
Entertainment & 0.03 & 0.01 & 0.67 & 0.43 & 0.62 & 0.10\\
Gold & 0.01 & 0.01 & 0.89 & 0.70 & 0.71 & 0.78\\
HealthScience & 0.51 & 0.72 & 0.37 & 0.43 & 0.47 & 0.38\\
Sports & 0.00 & 0.10 & 0.20 & 0.29 & 0.26 & 0.55\\
USA & 0.98 & 1.33 & \textbf{0.03} & 0.95 & 1.05 & 0.53\\
WeatherClimate & 0.74 & 0.58 & 0.56 & 1.48 & 1.13 & 0.26\\
\bottomrule
\end{tabular}
\label{tab:results_rq2}
\end{table*}

Table \ref{tab:results_rq2} presents $MSE_{\omega}^{man}$, $MSE_{\omega}^{woman}$, and the associated $p$-values from the bootstrapping procedure described in the previous section for Prompts 1 and 2. For abortion claims, both prompts resulted in highly statistically significant results ($\hat{D}_{\omega} = 1.3$ and $1.12$, with $p < 0.001$ for prompt 1 and 2 respectively), with the responses from LLMs being more aligned with men than women on this topic. For prompt 1, we observed statistically significant alignment of LLM responses with men's ratings for USA claims, and for prompt 2, we observed statistically significant alignment with men's ratings for Illegal Immigration claims. 

Overall, \texttt{GPT-3.5 Turbo} seems to perform relatively neutrally across a wide variety of topics when prompted without any mention of gender, as evidenced by relatively few statistically significant differences. However, in cases where  statistically significant differences in $MSE_{\omega}^{man}$ and $MSE_{\omega}^{woman}$ were observed, the neutrally-prompted \texttt{GPT-3.5 Turbo} model scores are more aligned with the men than women.  It is important to note that the alignment with men becomes particularly pronounced when evaluating claims about abortion. In such cases, women are especially pertinent as sources of evidence. This is because women are often the demographic most affected by the harmful consequences of misinformation related to abortion. Therefore, prioritizing claims from a male perspective can overlook the critical insights and experiences of women, who are directly impacted by these issues.

%% file: results_rq1.tex
\begin{table}[h]
\centering
\caption{Combined Results for Prompts 1 and 2}
\begin{tabular}{@{}lcccc@{}}
\toprule
Topic &  \(\hat{E}\) (Prompt 1) & Significance &  \(\hat{E}\) (Prompt 2) & Significance \\
\midrule
\multicolumn{5}{c}{Diverse Topics} \\
\midrule
Abortion & 0.56 & *** & 0.24 & *** \\
Black Americans & 0.63 & *** & 0.26 & ** \\
Illegal Immigration & 0.22 & ** & 0.05 & - \\
LGBTQ & 0.18 & * & 0.31 & *** \\
\midrule
\multicolumn{5}{c}{Other Topics} \\
\midrule
Entertainment & 0.30 & ** & 0.22 & * \\
Gold & 0.12 & *** & 0.10 & ** \\
HealthScience & 0.72 & *** & 0.34 & *** \\
Sports & -0.03 & - & -0.02 & - \\
USA & 0.56 & *** & 0.54 & *** \\
WeatherClimate & 0.47 & *** & 0.22 & ** \\
\bottomrule
\multicolumn{5}{c}{\footnotesize{Note: Significance levels are indicated as follows: \(*\) for p < 0.05, \(**\) for p < 0.01, and \(***\) for p < 0.001.}} \\
\end{tabular}
\label{tab:results_rq1}
\end{table}

%% file: discussion.tex
In this section, we discuss the implications and limitations of our findings, providing relevant insights to several stakeholders in the claim prioritization process.

\subsection{Implications for Fact-Checking Organizations}

Overstated differences in opinion by LLMs can lead to misguided priorities for fact-checking organizations seeking to use these technologies, raising concerns about procedural justice in their operations. Procedural justice emphasizes fairness in the processes that lead to outcomes, and in the context of claim prioritization, it pertains to representation and participation in the selection of claims for verification. Fact-checking organizations often have specific missions, such as combating misinformation affecting particular groups or topics. For instance, Univision explicitly states that their focus is fact-checking information that is relevant to the Spanish speaking community in the USA \cite{UnivisionIFCN2024}. In such cases, organizations using LLMs to assist in claim prioritization could use the perceived divergence of opinions between groups as a criterion for prioritizing new claims, asking questions like, ``Which issues are uniquely relevant to group X?'' However, an inaccurately high predicted disparity in opinions between group X and group Y could lead to misaligned priorities. This misalignment not only undermines the fairness of the fact-checking process but could also result in the organization, aiming to support group X, inadvertently focusing on less pertinent fact-checks. Consequently, members of group X might be exposed to a greater amount of misinformation on issues more relevant and impactful to them. Moreover, prompting the model without specifying a demographic could also yield answers that fail to reflect the views of relevant groups, as illustrated by the fact that in our experiments the neutral-prompted model is better aligned with men's views when asked about claims regarding abortion. The use of LLMs to prioritize misinformation should be used cautiously and tested meticulously through careful prompt engineering and fine-tuning to ensure procedural justice in the fact-checking pipeline.

On a practical note, when working with Google Fact-Check API, it became clear that claims related to U.S. politicians, the U.S. military, and U.S. foreign policy constituted a significant percentage of the English-language portion of the database. This is also reflected in our dataset, with roughly 25\% of claims being categorized in this topic. In our analysis, we observed a consistent pattern in the responses to both prompts: the model significantly overestimates the level of disagreement between men and women concerning the perceived group harm of claims related to this topic. Additionally, the responses generated by the model tend to exaggerate the level of perceived group harm when compared to human responses. This may be in part because the model associates political parties with groups likely to be harmed by these claims, while humans tend not to think of political parties as such groups. The data suggests that these topics represent the largest category of misinformation currently being prioritized for fact-checking in America, meaning the model may not accurately reflect the actual degree of gender-based disagreement in public perception on issues that are currently the focus of fact-checkers.

\subsection{Implications for LLM Developers}

The growing need for enhanced claim prioritization technologies is evident as fact-checkers face the herculean task of sifting through vast amounts of information to identify and prioritize information for fact-checking. In response to this demand, the integration of LLMs that can encapsulate and reflect diverse perspectives appears promising. Theoretically, LLMs trained on text data sourced from a rich variety of viewpoints have the potential to mirror these perspectives during inference, thereby adding context to the prioritization of claims in a way that is inclusive and representative of different societal groups’ opinions on what is deemed significant or harmful.

However, this potential is challenged by the presence of biases in the responses of LLMs, as examined in this research. In the first research question, we see that gender-conditioned prompts tend to exaggerate differences in opinion between men and women, and this is persistent in topics where we don't normally observe such differences. In the second research question, we find that LLMs prompted neutrally tend to better reflect men's opinion in the assessing of group harm related to abortion, an issue of critical importance in which women's perspectives are uniquely important and valuable.

The responsible development of applications of LLMs for claim prioritization requires a conscientious approach that acknowledges and addresses how societal factors, including gender, influence opinions and values. Developers should recognize these complex influences and steer clear of essentialist views that could lead to oversimplified and biased interpretations by the models. Those developing applications of LLMs for claim prioritization or subjective tasks must therefore engage in careful construction of training datasets, mindful prompting of models, and critical interpretation of model outputs. This approach is essential to ensure that LLMs serve as ethical, fair, and accurate tools in the fight against misinformation, representing the complexities of human opinion and societal dynamics without perpetuating or exacerbating biases.

The release of the \texttt{TopicMisinfo} dataset allows further research related to multiple dimensions of claim prioritization. While we hypothesize that this dimension of prioritization is the most likely to yield disagreements, further research should assess whether this exists along other dimensions as well. 

\subsection{Implications for Crowd-Sourcing Annotations}

In the challenging task of prioritizing claims for fact-checking, the variability and evolution of public opinions play a pivotal role. Given this dynamic landscape, the use of LLMs - or any AI model - for assisting in this process necessitates a rigorous and frequent recalibration to align with the changing societal views. Crowd-workers play an important role in the development of systems for claim prioritization. They provide a diverse, real-time perspective that is crucial for the calibration of LLMs, ensuring these models accurately reflect current public discourse and priorities. This ongoing recalibration, facilitated by crowd-workers, is particularly vital in the fact-checking domain, which is known for its rapid response to emerging information and trends. This approach is essential for maintaining the relevance and accuracy of LLMs in a field where opinions and discourse are in constant flux. 

\subsection{Limitations}

This paper presents valuable insights into the use of LLMs for claim prioritization in fact-checking pipelines, yet it is not without its limitations. One limitation of our study stems from the restricted diversity among the crowd-workers we employed. While our demographic survey contained the option for participants to identify their gender as non-binary, no participant self-identified this way. Consequently, our analysis was confined to examining the differences solely between men and women. This means that our study does not include insights into how other gender identities might compare, as we could not obtain data for these groups due to the homogeneity of our crowd-worker pool. Future research could benefit from exploring a broader array of gender perspectives. This expansion would provide a richer understanding of how LLMs handle diverse and complex gender identities.

Additionally, the paper's focus on a single LLM potentially limits the generalizability of the findings. Different LLMs, especially those trained with alternative methods or datasets, may exhibit varied behaviors and biases. Future studies should consider assessing a range of models to discern whether the observed amplification of gender biases is an inherent trait of this particular LLM or a broader characteristic of language models. 

Further, an increasing challenge when engaging with crowd-workers is that they may be using LLMs themselves to annotate data, as they often have financial incentives to increase their efficiency \cite{veselovsky2023artificial}. In the context of our research, if some workers did use LLMs, then our results likely understate the magnitude of the difference between LLMs and humans.  

Lastly, the paper and dataset is rooted in a perspective that focuses extensively on the United States, with topics in our dataset relevant to the culture of the United States and crowd-workers sourced from the United States. A more inclusive approach that considers diverse cultural contexts would be beneficial for developing LLMs that are truly representative of the global population.

%% file: appendix.tex
\input{poll_data}

\begin{table}[h]
\centering
\caption{Demographic analysis of \texttt{TopicMisinfo} dataset (\% of annotations labeled by group).}
\begin{tabular}{|l|l|}
\hline
\textbf{Category} & \textbf{Distribution} \\
\hline
Gender & 69\% Men / 31\% Women / 0\% Non-binary\\
\hline
Race & 66\% White / 25\% Asian / 2\% Black / 6\% NA \\
\hline
Sexual Orientation & 86\% Straight / 14\% Gay and Bisexual \\
\hline
\end{tabular}
\label{tab:demographics}
\end{table}

\begin{table}[h]
\centering
\caption{Claim Prioritization Dimensions Assessed in \texttt{TopicMisinfo} Dataset}
\begin{tabular}{|l|p{10cm}|}
\hline
\textbf{Operationalization} & \textbf{Description} \\
\hline
``Prioritization'' & Due to limited resources, fact-checking organizations must \emph{prioritize} which claims to check. In your opinion, should this claim be prioritized for fact-checking? (1 - 6 scale)\\
\hline
``General Public'' & To what extent will the claim be of interest to the general public? (1 - 6 scale)\\
\hline
``Group Harm'' &  How likely is this claim to disproportionately harm certain demographic groups more than others (e.g. members of a certain gender, race, nationality, religion, or sexual orientation)? (1 - 6 scale)\\
\hline
``Perceived Truth'' & Does the claim appear to be completely true, completely false, or contain both true and false information? (1 - 6 scale, higher meaning more false)\\
\hline
\end{tabular}
\label{tab:definitions}
\end{table}

\begin{algorithm*}
\caption{Bootstrapping procedure used in RQ2 for $H_0: \hat{D}_{\omega} = 0$ vs $H_1:  \hat{D}_{\omega} \neq 0$}
\label{alg:rq2}
\begin{algorithmic}[1]
    \STATE \textbf{Input}: \(dictionary\) of annotations for human and LLM by topic and claim, number of bootstrap iterations \( B \)
    \STATE \textbf{Output}: Bootstrap test statistics
    \STATE topic\_dict \(\gets\) dictionary containing counts per topic used to calculate $p$-values
    \FOR{\(b\) = 1 to \(B\)}
        \FOR{each topic in input \(dictionary\)}
            \STATE male\_scores, female\_scores, male'\_scores, female'\_scores, gpt\_scores \(\gets\) [], [], [], [], []
            \FOR{each claim in topic}
                \STATE m \(\gets\) male annotations for claim
                \STATE f \(\gets\) female annotations for claim
                \STATE g \(\gets\) gpt annotations for claim
                \STATE c \(\gets\) {concatenate}(m, f)\hfill\COMMENT{combine male and female into one sample}
                \STATE m' \(\gets\) sample(c, len(m)) \hfill\COMMENT{Sample of size m with replacement from c}
                \STATE f' \(\gets\) sample(c, len(f)) \hfill\COMMENT{Sample of size f with replacement from c}
                \STATE male\_scores \(\gets\) male\_scores \(\cup\) avg(m)
                \STATE female\_scores \(\gets\) female\_scores \(\cup\) avg(f)
                \STATE male'\_scores \(\gets\) male'\_scores \(\cup\) avg(m')
                \STATE female'\_scores \(\gets\) female\_scores \(\cup\) avg(f')
                \STATE gpt\_scores \(\gets\) gpt\_scores \(\cup\) avg(g)
            \STATE obs\_diff \(\gets\) MSE(male\_scores, gpt\_scores) - MSE(female\_scores, gpt\_scores) 
            \STATE diff \(\gets\) MSE(male'\_scores, gpt\_scores) - MSE(female'\_scores, gpt\_scores)
            \IF{diff \(\geq\) obs\_diff}
                \STATE topic\_dict[topic] += 1 \hfill\COMMENT{If we observe a significant difference in MSE}
            \ENDIF
            \ENDFOR
        \ENDFOR
    \ENDFOR
    \FOR{each topic in topic\_dict}
        \STATE topic\_dict[topic] /= \(B\) \hfill\COMMENT{topic\_dict stores $p$-values per topic}
    \ENDFOR
\end{algorithmic}
\end{algorithm*}

%% file: poll_data.tex
\begin{table*}[h]
\centering
\caption{Analysis of Gender-Based Differences in Opinion on Key Social Issues from the 2020 American National Election Studies \cite{ANES2021}, as emphasized by \citet{CAWP2023}. This comparison reveals that women tend to show greater sensitivity to the harms associated with LGBTQ rights, immigration policies, racial issues concerning African Americans, and abortion rights compared to men.}
\begin{tabular}{|>{\raggedleft\arraybackslash}p{0.35\textwidth}|p{0.15\textwidth}|p{0.15\textwidth}|p{0.15\textwidth}|}
\hline
\multicolumn{4}{|c|}{\emph{\begin{tabular}[c]{@{}c@{}}Should transgender people - that is, people who identify themselves as the sex or gender \\ different from the one they were born as - have to use the bathrooms of the gender they were \\ born as, or should they be allowed to use the bathrooms of their identified gender?\end{tabular}}} \\ \hline
\textbf{ANES 2020} & \textbf{Women} & \textbf{Men} & \textbf{Gender Difference} \\ \hline
Have to use the bathrooms of the gender they were born as & 44.7\% & 52.1\% & -7.4\% \\ \hline
Be allowed to use the bathrooms of their identified gender & 55.3\% & 47.9\% & 7.4\% \\ \hline \hline
\multicolumn{4}{|c|}{\emph{\begin{tabular}[c]{@{}c@{}}Which comes closest to your view about what government policy should be toward \\ unauthorized immigrants now living in the United States?\end{tabular}}} \\ \hline
\textbf{ANES 2020} & \textbf{Women} & \textbf{Men} & \textbf{Gender Difference} \\ \hline
Make all unauthorized immigrants felons and send them back to their home country & 12.3\% & 14.8\% & -2.5\% \\ \hline
Have a guest worker program that allows unauthorized immigrants to remain in the U.S. & 12.8\% & 15.8\% & -3.0\% \\ \hline
Allow unauthorized immigrants to remain in U.S. and to eventually qualify for U.S. citizenship  but only if they meet requirements & 55.9\% & 54.2\% & 1.7\% \\ \hline
Allow unauthorized immigrants to remain in U.S. \& eventually qualify for citizenship without penalties & 19.0\% & 15.3\% & 3.7\% \\ \hline \hline
\multicolumn{4}{|c|}{\emph{\begin{tabular}[c]{@{}c@{}}In general, does the federal government treat whites better than blacks, treat them both the same, \\ or treat blacks better than whites?\end{tabular}}} \\ \hline
\textbf{ANES 2020} & \textbf{Women} & \textbf{Men} & \textbf{Gender Difference} \\ \hline
Treat whites better & 50.3\% & 46.1\% & 4.2\% \\ \hline
Treat both the same & 37.9\% & 41.0\% & -3.1\% \\ \hline
Treat blacks better & 11.8\% & 12.9\% & -1.1\% \\ \hline \hline
\multicolumn{4}{|c|}{\emph{\begin{tabular}[c]{@{}c@{}}Would you be pleased, upset, or neither pleased nor upset if the Supreme Court reduced abortion rights?\end{tabular}}} \\ \hline
\textbf{ANES 2020} & \textbf{Women} & \textbf{Men} & \textbf{Gender Difference} \\ \hline
Pleased & 20.9\% & 22.9\% & -2.0\% \\ \hline
Upset & 51.4\% & 42.8\% & 8.6\% \\ \hline
Neither pleased nor upset & 27.7\% & 34.3\% & -6.6\% \\ \hline
\end{tabular}
\label{tab:poll_data}
\end{table*}